\documentclass[a4paper, fontsize=11pt]{article}

\renewenvironment{abstract}
               {\list{}{\rightmargin\leftmargin}%
                \item[\hspace*{1cm}\small\textbf{Abstract ---}]\relax}
               {\endlist}

\usepackage{authblk}

\usepackage[top=2.5cm, bottom=2.5cm, left=2.5cm, right=2.5cm]{geometry}
\usepackage[fleqn]{amsmath}
\usepackage{amssymb}
\usepackage{enumerate}
\usepackage{amsthm}
\usepackage[pdftex]{graphicx}
\usepackage{balance}
\usepackage{fancyhdr}
\usepackage{sidecap}
\usepackage{float}
\usepackage{scrextend}
\usepackage{changepage}
\usepackage{subfigure}
\usepackage{endnotes}
\usepackage{xcolor}
\let\footnote=\endnote

\newtheorem{Theorem}{Theorem}

\newtheorem{Example}[Theorem]{Example}

\newtheorem{Postulate}[Theorem]{Postulate}

\begin{document}

\title{\bf A paradox in the quantum-mechanical treatment of destructive measurements on photons}

\author[1]{Marcoen J.T.F. Cabbolet}
\author[2]{Yves Caudano}

\affil[1]{CLEA, Vrije Universiteit Brussel, Pleinlaan 2, 1050 Brussels, Belgium\footnote{email: Marcoen.Cabbolet@vub.be (corresponding author)}}
\affil[2]{Physics department, Namur Institute for Complex Systems (naXys) and Namur Institute of Structured Matter (NISM), Universit\'{e} de Namur, Rue de Bruxelles 61, 5000 Namur, Belgium\footnote{email: Yves.Caudano@unamur.be}}

\date{\today}

\maketitle
\vfill
\begin{abstract} 
Measurements on photons are frequently cited as confirmations of predictions of quantum mechanics (QM), in particular in the context of Bell's theorem. In this paper we show, however, that we cannot ever claim to have measured a property of a photon if we treat a destructive measurement {of the value of a property of a photon prepared in a superposition of eigenstates} in the framework of orthodox QM.
\end{abstract}

\section{Introduction}
The term `Bell's theorem' refers to a number of similar theorems establishing that, given a number of simple assumptions, quantum mechanics (QM) makes predictions about the outcome of measurements that cannot be reproduced by any local hidden-variable theory. In its original form, published by Bell \cite{Bell}, the theorem is about measuring properties of the subsystems of an entangled bipartite system prepared in a Bell state. In the decades that followed, experiments with photons have been done that are widely seen as empirical confirmations of those predictions of QM that cannot be reproduced by any local hidden-variable theory \cite{Freedman,Aspect,Zeilinger}. In the meantime an entire field called `quantum photonics' has arisen, which revolves around the use of photons to explore quantum phenomena.

The purpose of this paper is to show that measurements of properties of photons cannot be consistently treated in the framework of ``standard'' or ``orthodox'' QM (OQM), that is, QM in the widely accepted Copenhagen interpretation. Properties of photons that one might want to measure are, for example, its spin angular momentum $\vec{S}$, its frequency $\omega$, its energy $\hbar\omega$, its momentum $\hbar\vec{k}_0$, its wavelength $\lambda_0$, its orbital angular momentum $\vec{L}$, its total angular momentum $\vec{J} = \vec{L}+\vec{S}$, or its position $X$. Importantly, we do not question that properties of photons have been measured, nor do we question the good intentions of the experimenters: what we show is that the postulates of OQM do not ever allow us to say that a property of a photon has been measured {when a photon, prepared in a superposition of eigenstates of the operator $\hat{X}$, has been destroyed in a measurement of the value $x$ of the corresponding property $X$}. Consequently, not a single {one of these} destructive experiments on photons can be referred to as experimental support for OQM.
\newpage

\section{Identification of the paradox}\label{sect:paradox}

Physicists, as a rule, are used to `doing physics' without referring to the explicit formulation of the generally accepted postulates of the foundational theory in whose framework they are working. But when it comes down to QM, the situation is that there are no generally accepted postulates. It has to be thoroughly understood, however, that this situation has not arisen because it is not known what the postulates of QM are: this situation has arisen (i) because several versions of QM have been developed, each of which comes with its own set of postulates, and (ii) because there is no consensus about which version of QM should be used for `doing physics'. That being said, the most widely used version of QM is OQM, and its postulates have been explicitly formulated in \cite{Muller1,Muller2}: in the framework of OQM, it is precisely laid down in these postulates under which conditions we can say that a system `has' a property $X$ with value $x$. In fact, this is regulated by the Standard Property Postulate (SPP), {also called the eigenvalue-eigenstate link} \cite{Fine}, which can be traced back to works by Dirac {and Heisenberg} \cite{Dirac,Heisenberg}.

\begin{Postulate}[Standard Property Postulate (SPP)]\rm \ \\
A system `has' a property $X$ with quantitative value $x$ \textbf{if and only if} its quantum state is the eigenstate $| \ x \ \rangle$ of the associated operator $\hat{X}$. \hfill$\blacksquare$
\end{Postulate}

\noindent Often the SPP is tacitly assumed \cite{Muller3}, but it really is a postulate of OQM. Let us note that, in this work, we consider that a quantum state is represented by a vector $|\ \psi\ \rangle$ with norm $\langle \psi \vert \psi \rangle=1$, {and that an observable property $X$ is represented by a self-adjoined operator $\hat{X}$ with a discrete spectrum $\sigma(\hat{X})$, which we assume to be \emph{nondegenerate}}. The question is then how we can get a system in an eigenstate $| \ x_j \ \rangle$ of the operator $\hat{X}$ {when the system has been prepared in a pure quantum state $|\ \psi\ \rangle$ that is a superposition of such eigenstates, as in $|\ \psi\ \rangle = \sum_j \alpha_j| \ x_j \ \rangle$}. We find the answer in the Projection Postulate (PP), which traces back to works {by Dirac, von Neumann, and L\"{u}ders} \cite{Dirac,VonNeumann1,Luders}.

\begin{Postulate}[Projection Postulate (PP)]\rm\ \\
 \textbf{If} a measurement of the value of the property $X$ has been done on a system, prepared in a pure quantum state $|\ \psi\ \rangle$, with outcome $x_j \in \sigma(\hat{X})$, \textbf{then} immediately after the measurement the quantum state of the system is the eigenstate $| \ x_j \ \rangle$ of the associated operator $\hat{X}$.\hfill $\blacksquare$
\end{Postulate}

\noindent {A more general formulation of the PP is required to include cases where $\sigma(\hat{X})$ is continuous or degenerate \cite{Sudbery}; we note that the paradox remains in these cases.} Both the SPP and the PP are intimately interwoven with Born's Probability Postulate, which allows us to compute the probabilities of outcomes of measurements on a system prepared in a pure quantum state $|\ \psi\ \rangle$:
\begin{Postulate}[Born Rule]\rm\ \\
\textbf{If} a system with associated quantum state $|\ \psi\ \rangle$ is subjected to a measurement of the property $X$, represented by a maximal operator $\hat{X}$ with spectrum $\sigma(\hat{X}) = \{x_1, \ldots x_n\}$, \textbf{then} the probability $P^{|\ \psi\ \rangle}(x = x_j)\in [0,1]$ that the outcome $x$ is identical to $x_j \in \sigma(\hat{X})$ is given by the Born rule
\begin{equation}\label{eq:BornRule}
P^{|\ \psi\ \rangle}(x = x_j) = \left| \langle\ x_j\ |\ \psi\ \rangle\right|^2
\end{equation}
$\blacksquare$
\end{Postulate}
\noindent So, when calculating the probability of finding the value $x_j$ when measuring the property $X$ of a system prepared in the quantum state $|\ \psi\ \rangle$, we actually calculate the probability of a projection $|\ \psi\ \rangle \rightarrow |\ x_j\ \rangle$ as indicated by the occurrence of the term $\langle\ x_j\ |\ \psi\ \rangle$ in Eq. \eqref{eq:BornRule}. After all, by the SPP it is only in the eigenstate $|\ x_j\ \rangle$ of $\hat{X}$ that the system `has' property $X$ with value $x_j$.

Let us apply this to a photon, whose quantum state prior to a measurement of the value of its property $X$ is a superposition $|\ \psi\ \rangle$ of eigenstates of the operator $\hat{X}$: in this state it doesn't `have' the property $X$, as also remarked by Weinberg \cite{Weinberg}. So, to claim we have observed it having property $X$ with value $x_j \in \sigma(\hat{X})$, the photon must be in the eigenstate $|\ x_j\ \rangle$ of the associated operator $\hat{X}$, as dictated by the SPP. To get the transition from $|\ \psi\ \rangle$ to $|\ x_j\ \rangle$, we have to apply the PP. A paradox arises, however, when the act of measurement \emph{destroys} the photon. Immediately after the measurement, there is then no such thing as the quantum state of the photon: the PP then does not apply, hence the SPP does not apply. So, the paradox in the framework of OQM is that we can \textbf{never} claim that we have measured a value $x$ of a property $X$ of a photon that is destroyed in the measurement---if we claim that it had this property just prior to being destroyed, we make the error of tacitly applying the PP \emph{before} the measurement (for the SPP requires the photon to be in the eigenstate $|\ x\ \rangle$ to `have' the property).

\begin{Example}\rm

\begin{figure}[b!]
\centering
\hfill
\subfigure[experimental set-up]{\includegraphics[width=0.495\textwidth]{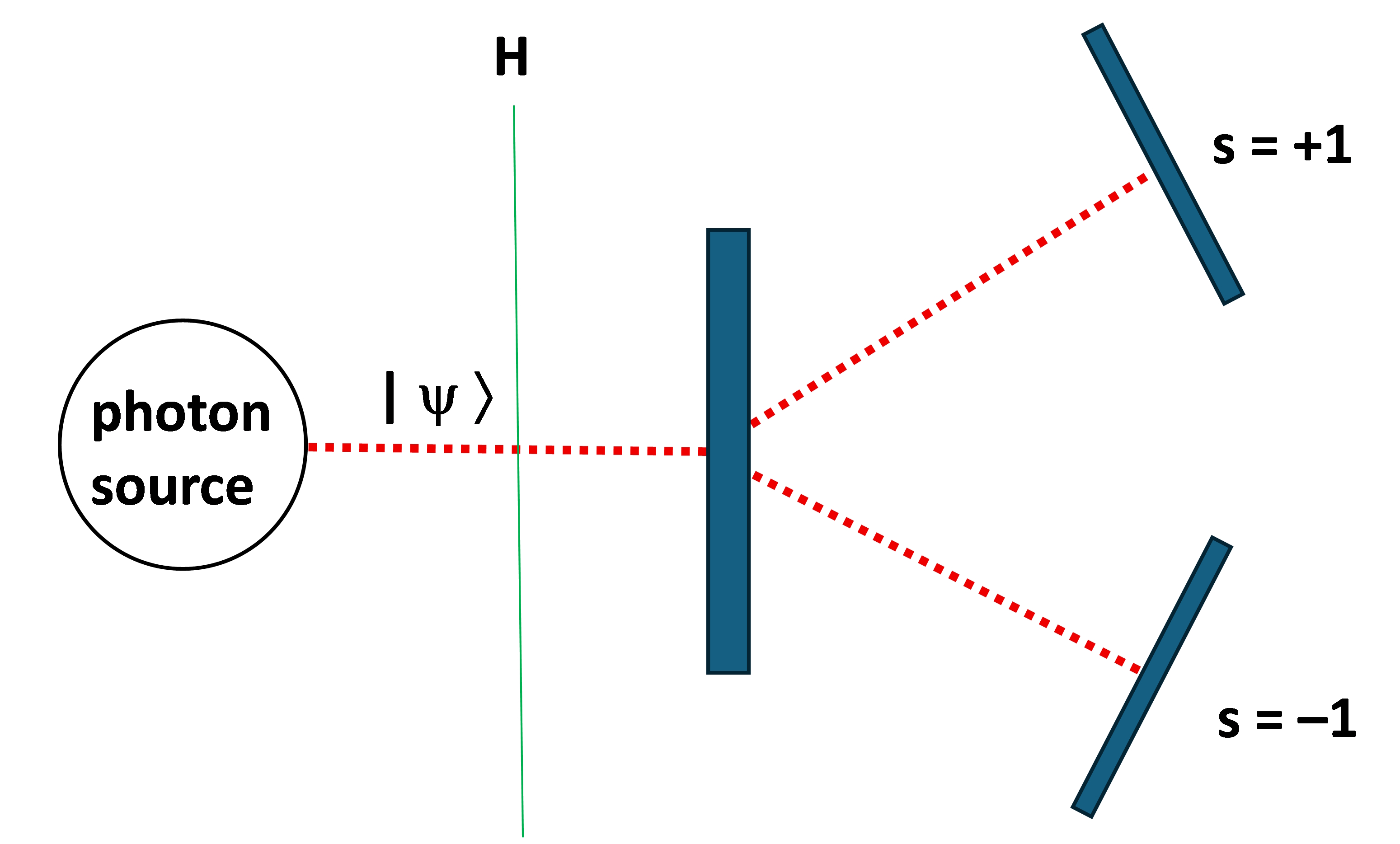}}
\hfill
\subfigure[scenario with outcome $s=+1$]{\includegraphics[width=0.495\textwidth]{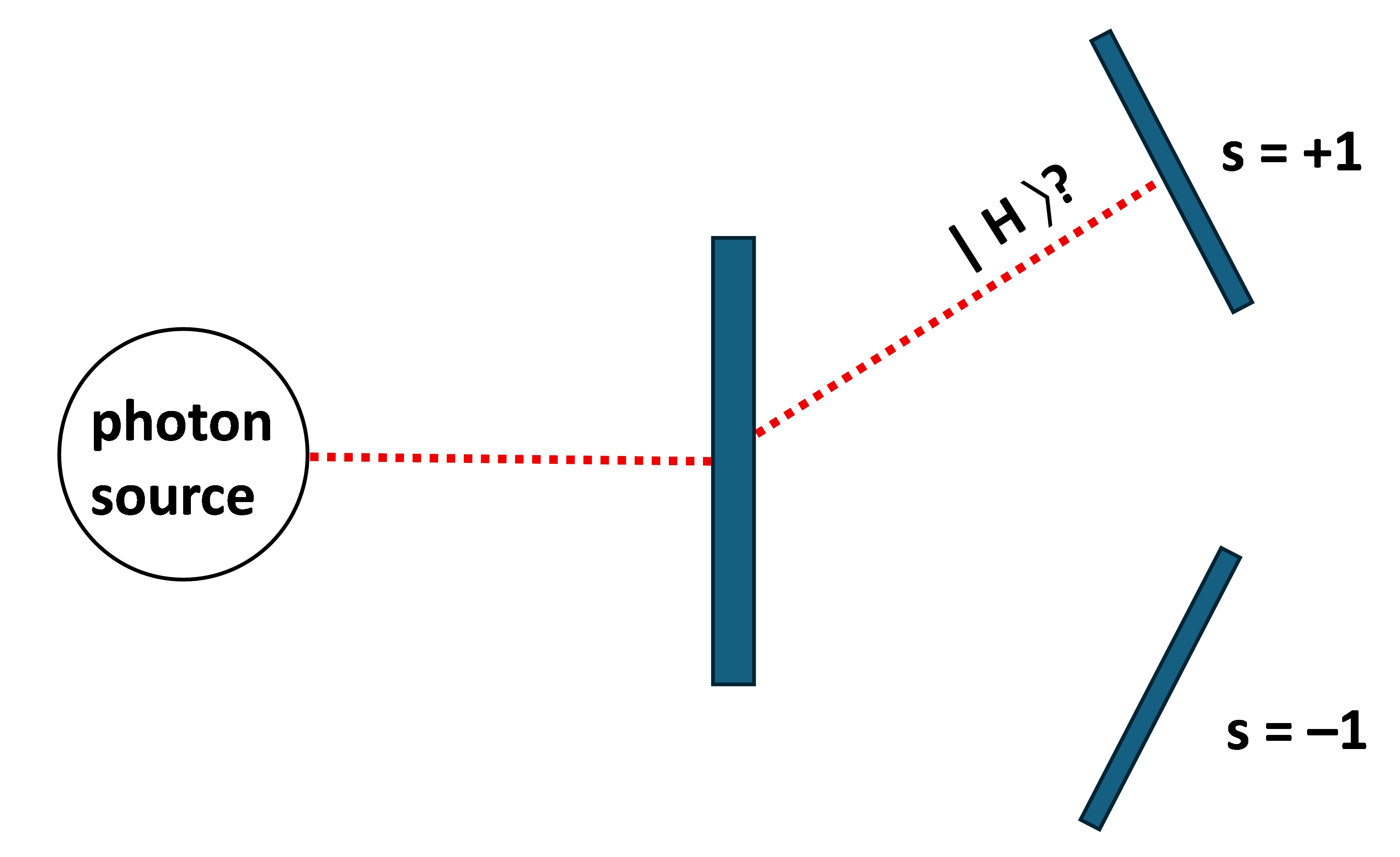}}
\hfill
\caption{Image (a) illustrates the experimental set-up, with Heisenberg cut $H$; image (b) illustrates the scenario when the photon has been detected by the upper detector.}
\label{fig:paradox}
\end{figure}

{Let $|\ H\ \rangle$ and $|\ V\ \rangle$ denote quantum states of a linearly polarized photon with respectively horizontal and vertical polarization; these are eigenstates of a spin operator $\hat{S}$, with
\begin{equation}\label{eq:eigenstates}
\left\{ \begin{array}{l}
\hat{S}|\ H\ \rangle = 1\cdot|\ H\ \rangle\\
\hat{S}|\ V\ \rangle = -1\cdot|\ V\ \rangle\
\end{array}\right.
\end{equation}
Now consider that a photon is prepared in a quantum state $|\ \psi\ \rangle$, for which
\begin{equation}\label{eq:PureState}
  |\ \psi\ \rangle = \frac{1}{\sqrt{2}}\left(|\ H\ \rangle + |\ V\ \rangle\right)
\end{equation}
We then measure the value of the polarization $s\in \{-1,+1\}$ by sending the photon into a measurement device, consisting of a two-channel polarizer followed by a photon detector as illustrated in Fig. \ref{fig:paradox}-(a), with the Heisenberg cut $H$ positioned between source and measurement device. Now suppose that we measure $s = +1$, as illustrated in Fig. \ref{fig:paradox}-(b). The paradox then occurs: for the photon to `have' the property $s=+1$ it must be in the eigenstate $|\ H\ \rangle$, but we have prepared it in the quantum state $|\ \psi\ \rangle$ of Eq. \eqref{eq:PureState}---note that the two-channel polarizer is part of the measurement device!---and the PP does not apply because the photon is destroyed in the measurement. So, there's our paradox: there is no way that we can claim that we have measured the property $s=+1$ of the photon, which we prepared in the quantum state $|\ \psi\ \rangle$.}

{Looking at Fig. \ref{fig:paradox}-(b), in hindsight one might be tempted to say that there is no paradox when we measure $s=+1$, because the photon was already in the eigenstate $|\ H\ \rangle$ after passing through the two-channel polarizer as indicated in Fig. \ref{fig:paradox}-(b): being in the eigenstate $|\ H\ \rangle$, the photon `has' the property $s=+1$ according to the SPP. So, no paradox! But then one is erroneously applying the PP before the measurement: projection takes place \emph{upon} measurement, and it doesn’t go backwards in time---prior to detection the photon was still in the quantum state $|\ \psi\ \rangle$.}\hfill$\blacksquare$
\end{Example}
\vfill

\section{Discussion}

While it has been noted before that photons are destroyed in measurement, e.g. in \cite{Sudbery}, the literature lacks a treatment of the paradoxical consequences thereof. Sects. \ref{sect:VacuumState}-\ref{sect:AdHoc} discuss three ways by which one may attempt to get out of this paradox:
\begin{enumerate}[(i)]
  \item we replace the quantum state of the photon by a (reduced) quantum state of the vacuum;
  \item we replace the PP by another projection postulate;
  \item we consider an \emph{ad hoc} modification of QM, postulating a two-step measurement process.
\end{enumerate}
Sect. \ref{sect:QNDM} briefly discusses quantum non-demolition measurements, Sect. \ref{sect:extra} discusses advanced measurement schemes and decoherence.

\subsection{Using a reduced quantum state of the vacuum}\label{sect:VacuumState}

Starting from the metaphysical presupposition that ``our'' photon was a part of the vacuum---we can view it as an excitation of the vacuum---we might consider a quantum state of the vacuum instead of the quantum state of a single photon: that way we avoid that the quantum state of the system vanishes upon measurement. Now the vacuum is an immensely complicated composite system: the average photon density due to CMB radiation is already some $4\times10^8\ m^{-3}$, so that might give an indication of the number of components of the vacuum state in a measuring device. So, only Laplace's demon might know the full quantum state of the vacuum. But for all practical purposes we can use a \emph{reduced} quantum state, {which we take to be a pure state $|\ n\ \rangle$ for some positive $n$ denoting the number of photons in the same quantum state $|\ \psi\ \rangle$ as ``our'' photon; here $|\ n\ \rangle$ is an eigenstate of the number operator $\hat{N}$ with $\hat{N}|\ n\ \rangle = n|\ n\ \rangle$, so if we measure the number of photons in the quantum state $|\ \psi\ \rangle$ then we are sure to get the outcome $n$ when the reduced state of the vacuum is $|\ n\ \rangle$. One might object to this treatment of the vacuum by pointing out that in practice there is always some ambiguity with respect to the photon number $n$: following the arguments in \cite{Sudbery}, the reduced state of the vacuum then should be a mixed state $\hat{\rho} = \sum_k \alpha_k |\ k\ \rangle\langle\ k\ |$ where $\alpha_k \in [0,1]$ is the probability that the photon number is $k$. We brush this objection aside: we acknowledge that it is true in general, but considering a mixed state makes things unnecessarily complex---to show the principle of the argument, it suffices to consider that the reduced quantum state of the vacuum is a pure state $|\ n\ \rangle$. So,}
\begin{enumerate}[(i)]
  \item prior to measurement, we associate to vacuum the (reduced) quantum state $|\ n\ \rangle$;
  \item upon measurement, we associate to vacuum the (reduced) quantum state $|\ n-1\ \rangle$.
\end{enumerate}
That way we formalize that by the measurement, the vacuum ends up with one photon less---we can view the measurement as a de-excitation of the vacuum. Since $\langle\ n-1\ |\ n-1\ \rangle=1$ we might think that this will resolve the paradox, but nothing is further from the truth: there are two problems that cannot be solved.

Firstly, applying the PP, if we have measured the value of a property $X$ with outcome $x$, then the quantum state $|\ n-1\ \rangle$, which we have associated to the vacuum upon measurement, is an eigenstate of the associated operator $\hat{X}$. The problem is then that there is no possible way to assign the measured property to the one photon whose properties we desperately want to measure: we have measured the value of a property of the vacuum, not of ``our'' photon.

Secondly, the two quantum states associated to the vacuum before and after the measurement are orthogonal: {$\langle\ n-1\ |\ n\ \rangle=0$}. Consequently, the Born rule as expressed by Eq. \eqref{eq:BornRule} doesn't apply for the calculation of probabilities. This rules out that the paradox will be resolved by using a reduced state of the vacuum: the paradox remains!

\subsection{Using an alternative projection postulate}\label{sect:NewPP}
Von Neumann already mentioned that the state of the measuring device has to be accounted for in a measurement \cite{VonNeumann1}. So, we may consider an alternative PP, in which the quantum state of the measuring device is taken into consideration: that way we also avoid that the quantum state vanishes upon measurement. We thus presuppose that for a property $X$ of a photon, represented by a maximal operator $\hat{X}$ with spectrum $\sigma(\hat{X}) = \{x_1, \ldots x_n\}$, there is a set $S$ of quantum states of the measuring device, $S = \{|\ \Phi_1\ \rangle, \ldots, |\ \Phi_n\ \rangle\}$, and a bijection $\phi: \sigma(\hat{X})\rightarrow S$, $\phi: x_k \mapsto |\ \Phi_k\ \rangle$, so that for every possible outcome $x_k \in \sigma(\hat{X})$ of a measurement of the value of the property $X$ there is a \emph{discernable} state $|\ \Phi_k\ \rangle \in S$ of the measuring device. (So, for each possible outcome $x_j$ there is, for example, a discernable position of a pointer on the device.) And each of the states $|\ \Phi_k\ \rangle \in S$ is then discernable from the quantum state $|\ \Phi_0\ \rangle$ that we associate to the measuring device \emph{before} it gives a reading. So, in practice, in experiments, the measurement device must be conceived such that
\begin{equation}\label{eq:device}
\langle\ \Phi_j\ |\ \Phi_k\ \rangle = \delta_{jk}
\end{equation}
where $\delta_{jk}$ is the Kronecker delta, so that we can distinguish the eigenvalues of the measured property. The alternative projection postulate that comes to mind is then the following:

\begin{Postulate}[Alternative Projection Postulate (APP)]\rm\ \\
\textbf{If} we measure the value of a property $X$ of a photon and we obtain the outcome $x_j$, \textbf{then} the tensor product of the quantum state $|\ \psi\ \rangle$ of the photon prior to measurement and the quantum state $|\ \Phi_0\ \rangle$ of the measuring device prior to measurement has changed discontinuously to the quantum state $|\ \Phi_k\ \rangle$ of the measuring device upon measurement:
\begin{equation}\label{eq:APP}
|\ \psi\ \rangle \otimes|\ \Phi_0\ \rangle \rightarrow |\ \Phi_k\ \rangle
\end{equation}
$\blacksquare$
\end{Postulate}

\noindent {The PP by von Neumann, in which the measuring device is taken to be a quantum system, has been critically discussed in \cite{Sudbery}: the above APP is not the same as that PP by von Neumann. More specifically, the above APP goes with von Neumann in that it associates a pure state---in our case: $|\ \psi\ \rangle \otimes|\ \Phi_0\ \rangle$---to the combined system prior to measurement. But von Neumann puts the Heisenberg cut \emph{inside} the observer, between the optic nerve and the brain \cite{VonNeumann1} (p. 439). The vacuum between the device and observer is then part of the quantum system, and so are the retina and the image on the retina obtained by looking at the device: it is then the consciousness of the observer that is responsible for the collapse of the wave function. The above APP disagrees with that positioning of the Heisenberg cut: we put the Heisenberg cut between the combined system and the observer. By looking, the observer ``measures'' the position of the pointer---by looking there's an effect of the device on the observer but not the other way around---so there simply is a quantum state prior to measurement (looking, that is) and a quantum state upon measurement: that's what the APP expresses. And it's unambiguous, because the different positions of the pointer are unambiguous.}

By accepting this APP instead of the ``standard'' PP as a postulate of OQM, we avoid the earlier situation that we have a projection postulate that does not apply to a single photon. But we are still not out of the woods, far from it even. For if we use this APP, we no longer consider the probability of a transition $|\ \psi\ \rangle \rightarrow |\ x_j\ \rangle$, which underlies the occurrence of the inner product $\langle\ x_j\ |\ \psi\ \rangle$ in Eq. \eqref{eq:BornRule}: we now consider the probability of the  transition $|\ \psi\ \rangle \otimes|\ \Phi_0\ \rangle \rightarrow |\ \Phi_k\ \rangle$ in Eq. \eqref{eq:APP}. The problem, however, is that the quantum state $|\ \Phi_0\ \rangle$, which we associate to the measuring device prior to measurement, is by Eq. \eqref{eq:device} assumed to be orthogonal to the quantum state $|\ \Phi_k\ \rangle$, which we associate to the measuring device prior to measurement. So, we run into the same problem as discussed in Sect. \ref{sect:VacuumState}: the Born rule, as expressed by Eq. \eqref{eq:BornRule}, fails for the calculation of the probability that a measurement of the value of the property $X$ of the photon has the outcome $x=x_j$.

Even if we manage to fix this problem, our formalism \emph{still} leaves no room to link the quantum state $|\ \Phi_k\ \rangle$ of the measuring device to a property of the photon. That is to say, merely finding the measuring device in the state $|\ \Phi_k\ \rangle$ doesn't allow us to make any assertion about a property of a photon that existed in the device just prior to measurement: we cannot apply the SPP! In other words: if we want to use the above APP, and we want to link the state of the device upon measurement to a property of a photon that was destroyed by the measurement, then we force ourselves to rewriting the postulates of QM: this inevitably results in yet another version of QM that cannot be called OQM!

\subsection{Ad hoc modification}\label{sect:AdHoc}

A perhaps more fruitful way to resolve the paradox is to replace the standard PP by a new Projection Postulate, by which a destructive measurement on photons becomes a process consisting of two steps. So, let us again {put the Heisenberg cut between the device and the observer as in Sect. \ref{sect:NewPP}, and} consider the tensor product $|\ \psi\ \rangle \otimes|\ \Phi_0\ \rangle$ as the initial state, where the quantum state of the photon is a pure state $|\ \psi\ \rangle =\sum_k \alpha_k |\ e_k\ \rangle$ in a Hilbert space spanned by an orthonormal basis $\{|\ e_1\ \rangle, \ldots, |\ e_n\ \rangle\}$, and let us postulate the following with $|\ 0\ \rangle$ denoting the no-photon-state:
\begin{enumerate}[(i)]
  \item the system-meter interaction first yields a transition $|\ \psi\ \rangle \otimes|\ \Phi_0\ \rangle\rightarrow \sum_k\alpha_k |\ e_k\ \rangle\otimes|\ \Phi_k\ \rangle$;
  \item immediately thereafter the projection $\sum_k\alpha_k |\ e_k\ \rangle\otimes|\ \Phi_k\ \rangle\rightarrow |\ 0\ \rangle\otimes|\ \Phi_k\ \rangle$ takes place.
\end{enumerate}
Let us note that, by repeating the system-meter interaction many times and performing quantum state tomography, it is in principle provable experimentally that the interaction produces the reduced mixed state of the meter represented by the state operator $\sum_k|\alpha_k|^2 |\ \Phi_k\ \rangle\langle\ \Phi_k\ |$. Then the projective measurement of the meter state provides now both the correct probabilities $|\alpha^k|^2$ and the appropriate final meter state $|\ \Phi_k\ \rangle\langle\ \Phi_k\ |$.

Having observed the final meter state, the preceding entangled state $\sum_k \alpha_k |\ e_k\ \rangle\otimes|\ \Phi_k\ \rangle$ then allows us to do statements about the property of the photon. At first glance this ad hoc modification of the postulates of QM may seem a way out of the paradox, but we have to understand that we have not solved the paradox within the framework of OQM: we have merely created another (unorthodox) version of QM by replacing the standard PP with a ``two-step PP''. In OQM, the paradox remains.

\subsection{Quantum non-demolition measurements}\label{sect:QNDM}

\begin{SCfigure}[1][b!]
\includegraphics[width=0.5\textwidth]{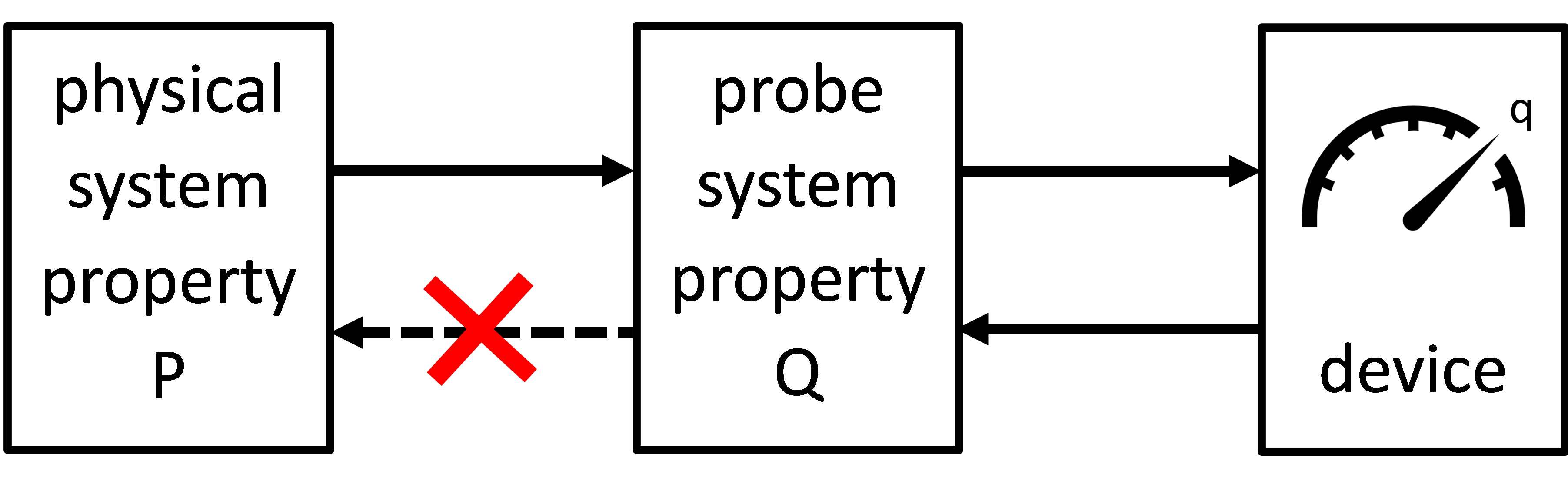}
\caption{Schematic illustration of a quantum non-demolition measurement. Upon measurement, there is no back action from the probe system to the physical system.}
\label{fig:QNDM}
\end{SCfigure}

In recent decades a whole area of research has evolved around so-called quantum non-demolition measurements, first discussed by Braginsky \emph{et al}. \cite{Braginsky}. The concept has been successfully applied to measure the number of photons in a cavity without destroying them \cite{Haroche}. In a quantum non-demolition measurement we have a physical system whose property $P$ we want to measure, and a `probe system' whose property $Q$ we actually measure. The idea is that there is an interaction between the physical system and the probe system, such that the measured value $q$ of the property $Q$ of the probe system is caused by the value $p$ of the property $P$ of the physical system, but such that the measurement itself of the value of the property $Q$ of the probe system causes no back action towards the physical system. See Fig \ref{fig:QNDM} for an illustration.

The idea is thus that this offers a way out of the paradox: with a quantum non-demolition measurement we might be able in the future to measure any property $P$ of a photon without destroying it. That latter statement may be true, but the paradox is about \emph{destructive} measurements on photons: that paradox remains, unabated---at best, the paradox becomes obsolete when destructive measurements on photons become obsolete. As a side note, quantum non-demolition measurements may be susceptible to the general criticism on measurements in physics that often an \emph{inference to the best explanation} is passed off as an \emph{observation} \cite{Cabbolet2}; a discussion thereof is outside the scope of this paper.

\subsection{{Advanced measurement schemes and decoherence}}\label{sect:extra}

{In the course of this paper, we focused on the simplest case of the standard projective measurement of a photon prepared in pure state, as it highlights the essence of the paradox. Of course, since the inception of quantum mechanics, physicists developed more complex measurement protocols and models, either to extract ever more specific information on the studied quantum systems, or to improve our understanding of the quantum measurement problem. In the following, we discuss the paradox in the context of more advanced measurement schemes and models, going beyond the textbook case addressed so far. In particular, we will consider the measurement of mixed states, the general framework of measurement known in quantum information theory as positive operator valued measures (POVM), and the inclusion of decoherence. We will take these opportunities to further illustrate the paradox through simple examples.}

\subsubsection{{Mixed states}}

{For mixed states, projective measurements may sometimes reveal preexisting properties, so that the paradox need not occur systematically. However, in all generality, it still stands. Here, we first consider the typical case where destructive measurements are fine. Then, we illustrate the fundametal issues at hand when considering destructive measurements of photons in a mixed state.}

{Suppose a photon is prepared with probability $\lambda_i$ in the eigenstate $|\ x_i\ \rangle$ of an operator $\hat{X}$ with spectrum $\sigma(\hat{X}) = \{x_1, \ldots x_n\}$. Due to our probabilistic knowledge of the state in which the photon was prepared,}
{ we associate a \emph{mixed state} $\hat\rho$ to the photon, described by the density operator
\begin{equation}\label{eq:MixedState}
  \hat\rho = \sum_{i=1}^n \lambda_i |\ x_i\ \rangle\langle\ x_i\ |
\end{equation}
with $0\leq \lambda_i\leq 1$ and $\sum_{i=1}^{n} \lambda_i = 1$. Importantly, the probabilities $\lambda_i$ in Eq. \eqref{eq:MixedState} reflect our ignorance: the photon is actually in one of the eigenstates $|\ x_i\ \rangle$, but we don't know which one it is. It is merely the case that when we repeat this preparation, a large number of times, then in a fraction of cases $\lambda_1$ the photon is in the eigenstate $|\ x_1\ \rangle$, in a fraction of cases $\lambda_2$ the photon is in the eigenstate $|\ x_2\ \rangle$, etc. Being in an eigenstate $|\ x_i\ \rangle$ means that the photon actually has the property $X$ with a definite value $x_i \in \sigma(\hat{X})$ \emph{prior to measuring it}: by a} {projective measurement of the observable $\hat{X}$,} {we find out which value $x_i$ that is, and it is in that case not important that the photon is destroyed in the measurement---for it already had the property $X$ with that value $x_i$ prior to measurement. So, in this case, the paradox does not occur: for a photon in a mixed state $\hat\rho$ of Eq. \eqref{eq:MixedState}} {resulting from a random preparation of one of the eigenstates of the operator associated to the property $X$,} {we can actually use a destructive measurement to determine what the value is of the property $X$ that the photon `has' prior to measurement.}

{However, not all mixed states result from a random preparation of solely the eigenstates associated to the measured property $X$. Let us consider two incompatible observables $X$ and $Z$, for example given by the Pauli matrices $\hat\sigma_x$ and $\hat\sigma_z$ for the sake of simplicity. Their respective eigenstates $\{\vert D \rangle, \vert A \rangle\}$ and $\{\vert H \rangle, \vert V \rangle\}$ are associated to the diagonal, anti-diagonal, horizontal, and vertical polarisations, respectively. Now, let us take a random, equiprobable preparation of the states $\vert D \rangle$ and $\vert H \rangle$. The associated density operator is thus
\begin{equation}\label{eq:MixedState2}
\hat\rho=\frac{1}{2} \vert D\rangle\langle D\vert + \frac{1}{2} \vert H\rangle\langle H\vert = \frac{1}{4}
\begin{pmatrix}
3 & 1 \\
1 & 1
\end{pmatrix},
\end{equation}
where the matrix is expressed in the basis of $\hat\sigma_z$. In this case, a measurement of the property $Z$ cannot be said to generally reveal a property existing prior to measurement because whenever the system measured is in the state $\vert D\rangle=\frac{1}{\sqrt{2}}(\vert H\rangle + \vert V\rangle )$, it corresponds to a quantum superposition of the eigenstates of the property $Z$. Each time the detector placed behind the horizontal-polarization output port of the polarizing beam splitter registers an event, there is a probability $\frac{1}{4}$ that the state of the photon prior to measurement was $\vert D \rangle$, and thus in a superposition of the eigenstates of $\hat\sigma_z$. A similar reasoning goes for a measurement of the property $X$, since $\vert H\rangle=\frac{1}{\sqrt{2}}(\vert D\rangle + \vert A\rangle )$. This shows that the paradox remains with general mixed states because it} {occurs when we use a destructive measurement to determine the value of the property $X$ of a photon prepared in a quantum state} {that involves} {a superposition $|\ \psi\ \rangle=\sum_i \alpha_i |\ x_i\ \rangle$ of eigenstates of the corresponding operator $\hat{X}$,} {and mixed states may still involve superpositions.}

{Let us also stress the following issue arising from the use of mixed states. While a given state preparation procedure yields a single density operator, the converse is not true: a given density operator corresponds to many possible mixtures involving different states. Therefore, given a mixed state and a property X such that both associated operators $\hat{\rho}$ and $\hat{X}$ are diagonal in the same basis, one could not generally claim that a measurement of $X$ would reveal a pre-existing value of the property $X$. As an illustration, we can use the previous example, where the density operator (\ref{eq:MixedState2}) is diagonal in the (unnormalized) basis $\{\vert H\rangle + \vert D\rangle, \vert V\rangle - \vert A\rangle\}$. In view of the preparation procedure described before, which generates only $\vert D\rangle$ and $\vert H\rangle$ states, a measurement of the linear polarization $P$ in the $\{\vert H\rangle + \vert D\rangle, \vert V\rangle - \vert A\rangle\}$ basis could not be deemed as revealing pre-existing values of the property $P$ for states prepared in eigenstates of the incompatible properties $X$ or $Z$. Another striking illustration is the case of fully non-polarized light, whose density operator is given by one half the identity matrix. It is diagonal in every polarization measurement basis: taking an ignorance interpretation of this density operator (as would be standard in classical physics), i. e. considering that the actual polarization state of each photon is perfectly random, one cannot consider that all measurements of an arbitrary polarization property $P$ would systematically reveal a pre-existing value of $P$.}

\subsubsection{{Advanced measurement schemes}}
{In order to improve the modelling of quantum experiments, more advanced measurement protocols were devised, especially in the context of quantum information theory. A very general and widely used measurement scheme that goes beyond projective measurements is notably that of positive operator valued measures (POVM) (see for example \cite{Nielsen}). While such advanced schemes improve substantially the description of possible information flows in quantum experiments, including in non-ideal and noisy conditions, they typically do not deal explicitly with the quantum-to-classical frontier. In other words, the randomness of quantum measurements still fundamentally results from the same two-step process than the one used in orthodox QM for projective measurements (unitary evolution followed by a random, non-unitary outcome). As a result, while they essentially refine the computation of probabilities, they remain dependent of putting an \emph{ad hoc} Heisenberg cut between the quantum and classical worlds and, thus, remain affected by the paradox. (Note that we will discuss the decoherence approach later on.)}

{As an illustration, let us look at the issues arising from the consideration of a POVM measurement \cite{Nielsen} involving $m$ non-orthogonal measurement operators $\hat{E_i}$, which are positive semi-definite and verify the completeness relation $\sum_{i=1}^m\hat{E_i}=\hat{I}$ (where $m>n$, with $n$ the dimension of the Hilbert space). Given the density operator $\hat\rho_0$ of the input quantum state, the measurement outcome $i$ is then obtained with probability $P(i\vert0)=\mathrm{tr} \hat{\rho}_0\hat{E}_i$. This is sufficient to compute the probabilities of the measurement outcomes but the definition actually does not specify what is measured. In particular, at this level of definition of the POVM, the final state of the system is unknown, so that we cannot state its actual properties according to orthodox QM. In order to determine the post-measurement state, one needs to define a set of $m$ Kraus operators $\hat{K}_i$ such that $\hat{E_i}=\hat{K}_i^\dagger\hat{K}_i$. Now, assuming the outcome $i$ was obtained, the post-measurement state of the quantum system is given by
\begin{equation}
\hat{\rho_i}=\frac{\hat{K}_i \hat{\rho}_0 \hat{K}_i^\dagger}{\mathrm{tr}\hat{\rho}_0\hat{E}_i} \quad\quad\mathrm{or}\quad\quad \vert\psi_i\rangle=\frac{\hat{K}_i \vert \psi_0\rangle}{\sqrt{\langle \psi_0\vert\hat{E}_i\vert\psi_0\rangle}},
\end{equation}
where the second equation is valid when the initial state is a pure state ($\hat\rho_0=\vert\psi_0\rangle\langle\psi_0\vert$). Note that the set of Kraus operators is not unique, as they are defined up to multiplication by an arbitrary unitary operator. In an experimental context, they depend on the specificity of the practical implementation of the POVM $\{\hat{E}_i\}$. Now, we can see that the POVM measurement scheme does not yield an immediate interpretation of its outcomes in terms of measuring the value of a property $X$ of an arbitrary superposition $\vert\psi_0\rangle=\sum_k \alpha_k \vert x_k\rangle$, in the sense of postulate 2 of orthodox QM, for several reasons. First, the possible final states depend on the specific implementation of the operators $\{\hat{E}_i\}$. Second, the possible final states are not orthogonal, so that they do not distinguish states according to different values of a given property. Third, the non-orthogonal outcomes also imply that the outcomes of a repeated POVM measurement are not reproducible. To specifically address the case of destructive measurement of photons, we consider Naimark's dilation theorem: to build a POVM, one needs in practice to add an ancillary system to the quantum system of interest and perform a standard projective measurement on the larger Hilbert space built on the two subsystems. The non-unicity of the decomposition $\hat{E_i}=\hat{K}_i^\dagger\hat{K}_i$ results from the multiple possible choices of the projective measurement that yields the POVM $\{\hat{E}_i\}$. Now, this larger projective measurement on the joint photon and ancillary system is susceptible to all the troubles we already identified within orthodox QM when the photon is destroyed in the measurement process. In particular the Hilbert space dimension is not preserved and one cannot repeat the measurement. This shows that, with all the power that modern approaches to measurement may bring pragmatically, they are no less susceptible to the foundational issues discussed here.}

{For specific input states, some outcomes of destructive POVM measurement performed on photons may nevertheless be interpreted straightforwardly within the framework of orthodox QM. For example, for the equiprobable mixture of $\vert D \rangle$ and $\vert H \rangle$ described by (\ref{eq:MixedState2}), the POVM built around the three measurement operators $\{\hat{E}_1=\frac{1}{1+\vert\langle D \vert H \rangle\vert}\vert A\rangle\langle A\vert,\hat{E}_2=\frac{1}{1+\vert\langle D \vert H \rangle\vert}\vert V\rangle\langle V\vert, \hat{E}_3=\hat{I}-\hat{E}_2-\hat{E}_1 \}$ performs optimal state discrimination between $\vert D \rangle$ and $\vert H \rangle$ \cite{Nielsen,Barnett}. Whenever, the outcome 1 is obtained, we know that the initial state could not be $\vert D \rangle$, so that it must have been $\vert H \rangle$ and the photon prior to measurement had the value $H$ of the property $Z$. Similarly, the outcome 2 determines that the photon had the value $D$ of the property $X$. These inferences are permissible because we know that the quantum states composing the mixture had those properties prior to measurement, as the uncertainty about them was of a strictly classical nature. Outcome 3, on the other hand, is susceptible to the paradox because the state $\vert D \rangle$ and $\vert H \rangle$ correspond to superpositions in the basis in which (\ref{eq:MixedState2}) is diagonal (which is related to $\hat{E}_3$). Let us note that, even if the considered POVM measurement were not destructive, the measurement process yielding outcomes 1 and 2 would nevertheless modify the photon quantum states. This non-ideal measurement can thus  determines appropriately the pre-existing property some of the time but always modifies it in the process.}

{As another illustration, we can consider post-selected weak measurements \cite{Aharonov}, which have attracted much interest over the last decades for both fundamental and practical reasons (see for example \cite{Dressel} for an introductory review). In a typical optical weak measurement, a photon is prepared in an initial polarization state that is an eigenstate $\vert \psi_i\rangle=\alpha_H \vert H\rangle + \alpha_V \vert V\rangle$ of the observable $\hat\sigma_i$, with a transverse spatial profile given by a Gaussian $G(x)$. The polarization Hilbert space defines the observables of interest, while the spatial profile plays the role of the meter in a von Neumann type of measurement.  The polarization and position degrees-of-freedom of the photon then interact weakly when it propagates through a thin birefringent crystal, causing two orthogonal polarizations, say $\vert H\rangle$ and $\vert V \rangle$, to propagate in slightly different directions.  The latter interaction can be represented by an effective Hamiltonian $H_{int}= \delta(t)\, x_0\, \hat{\sigma}_z \otimes \hat{p}_x$ (with $\hat{p}_x$ the momentum operator). As a result of the weak interaction, the photon state has now become $\alpha_H G(x-x_0) \vert H\rangle + \alpha_V G(x+x_0) \vert V\rangle$, where the shifts $\pm x_0$ are assumed much smaller than the width $w_0$ of the Gaussian. Their values correspond to the product of the eigenvalues of $\hat\sigma_z$ with the parameter $x_0$, which characterizes the strength of the interaction. The last step, post-selection, corresponds to a projective measurement of the observable $\hat\sigma_f$, where one keeps only the measurement results associated to the post-selected eigenstate $\vert \phi_f\rangle=\beta_H \vert H\rangle + \beta_V \vert V\rangle$. After successfull post-selection, the final state becomes approximately $G(x-x_0\langle \hat\sigma_z\rangle_w) \vert\phi_f\rangle$, where $\langle \hat\sigma_z\rangle_w=\langle \phi_f \vert \hat\sigma_z\vert\psi_i\rangle/\langle \phi_f \vert \psi_i\rangle$ is a complex number called the weak value of the observable $\hat\sigma_z$. We see that it plays a role similar to the eigenvalues of $\hat\sigma_z$ before post-selection since the (now complex) shift of the Gaussian wavepacket is given by the product of the weak value with the parameter $x_0$. The interpretation of weak values has led to many debates and it is certainly not our goal to dwell on those aspects here. Nevertheless, one proposed interpretation of the weak value is that its real part could represent the value of a property ($Z$ in this case) of the pre- and post-selected ensemble \cite{Aharonov} (i.e. the subset of photons which had initially the value $i$ associated to the eigenstate $\vert \psi_i\rangle$ of the property $I$ represented by $\hat\sigma_i$ after pre-selection, which ended up with the value $f$ associated to the eigenstate $\vert \phi_f\rangle$ of the property $F$ represented by $\hat\sigma_f$ after post-selection). While such a bold interpretation is obviously in conflict with the standard eigenvalue-property link described by postulate 1, a more palatable interpretation along those lines in the context of strictly orthodox QM is that the real part of the weak value represents the best conditional estimate of the value of the property $Z$ given our knowledge of the initial and final states of the photon polarization in the course of the experiment \cite{Hall, Dressel2}. Here, we would like to point out that any destructive post-selection measurement actually does not allow to even define properly the pre- and post-selected ensemble (at a purely formal level) since we cannot claim, according to orthodox QM, that the photon ever had the value $f$ associated to the eigenstate $\vert \phi_f\rangle$ of the property $F$ represented by $\hat\sigma_f$. Needless to say, this paradox doesn't preclude the usefulness of weak measurements for all practical purposes \cite{Xu,Ballesteros}!}

\subsubsection{{Decoherence}}
{Much effort has been devoted to circumvent the arbitrary two-step measurement process formalized by the standard postulates of QM, i.e. a deterministic unitary evolution followed by a random non-unitary projection. A preeminent attempt at solving ``the measurement problem", as it is often referred to, revolves around the notion of decoherence \cite{Zurek1} and related approaches developed subsequently, such as quantum darwinism \cite{Zurek2,Zurek3}. While these advances brought much progress towards explaining why the world could look classical while being quantum, they do not yet explain how a single classical outcome is selected among all the possible ones. Therefore, today's formalism of decoherence does not solve the paradox pinpointed by our paper, as we now explain in more details.}

{Let us first consider the simple von Neumann measurement of the photon polarization using a birefringent plate. As we just saw when dealing with weak measurements, the photon state after the plate is given by $\vert\Psi\rangle=\int [\alpha_H G(x-x_0) \vert H\rangle+\alpha_V G(x+x_0) \vert V\rangle]\otimes \vert x\rangle dx$ (where, now, $x_0$ is not necessarily small with respect to the width $w_0$ of the Gaussian). In such a state, the probability of finding the photon at the position $x$ is given by $\vert \alpha_H\vert^2 \vert G(x-x_0)\vert^2+\vert \alpha_V\vert^2 \vert G(x+x_0)\vert^2$. We observe that the two shifted wavepackets never interfere, even when they overlap, because they are entangled with orthogonal polarization states of the photon. However, interferences can still be recovered, in particular by projecting the photon on a specific polarization state.\textsuperscript{1}}\footnotetext[1]{{In the weak measurement scheme described previously, the role of post-selection is precisely to enable interference between the two fully overlapping wavepackets by projecting them on a common polarization eigenstate $\vert \phi_f\rangle$. This leads effectively to a single wavepacket with average position related to the real part of the weak value and average momentum related to the imaginary part of the weak value (to first order of approximation).}}
{Besides the Hilbert space $\mathcal{H}_S\oplus\mathcal{H}_M$ of the }{ photon (defining the system and meter in the von Neumann model of measurement), we now consider the Hilbert spaces of the macroscopic detector $\mathcal{H_D}$ and its even larger environment $\mathcal{H_E}$ as well. For long interaction times and a very large environment, the decoherence approach would ascribe to the complete photon-detector-environment system a state analogous to:
\begin{equation}\label{eq:decoherence1}
 \int [\alpha_H G(x-x_0) \vert H\rangle\otimes \vert x\rangle\otimes \vert \mathcal{D}(x)\rangle\otimes \vert \mathcal{E}(x)\rangle+\alpha_V G(x+x_0) \vert V\rangle\otimes \vert x\rangle\otimes \vert \mathcal{D}(x)\rangle\otimes \vert \mathcal{E}(x)\rangle] dx,
\end{equation}
where $\vert\mathcal{D}(x)\rangle$ represents the quantum state of the detector if it detects the photon at position $x$ and $\vert\mathcal{E}(x)\rangle$ represents the resulting quantum state of the environment when the detector locates the photon at position $x$. In such account of decoherence, the states of the environment would end up being orthogonal for different position measurement of the photon: $\langle \mathcal{E}(x) \vert \mathcal{E}(x^\prime)\rangle\approx 0$ if $x\ne x^\prime$. Thus, the macroscopic states of the detector cannot interfere anymore because they are entangled with orthogonal states of the environment. Of course, a truly precise account of the detection should also explain quantitatively why the position basis would be preferred in this process (i.e. it would justify why it is the position eigenstates that end up entangled with orthogonal eigenstates of the environment), but one clue is that the interactions are typically distance dependent and favour thus localized states \cite{Zurek3}. Because the full environment state is unknowable, we must evaluate the photon-detector state by tracing out the environment. This leads to the following density operator, diagonal in the position basis:
\begin{eqnarray}\label{eq:decoherence2}
 \rho_{SM\mathcal{D}}=\int \vert\alpha_H\vert^2 \vert G(x-x_0)\vert^2 \vert H\rangle\langle H\vert\otimes \vert x\rangle\langle x\vert\otimes \vert \mathcal{D}(x)\rangle\langle\mathcal{D}(x)\vert dx \nonumber\\
+ \int \vert\alpha_V\vert^2 \vert G(x+x_0)\vert^2 \vert V\rangle\langle V\vert\otimes \vert x\rangle\langle x\vert\otimes \vert \mathcal{D}(x)\rangle\langle\mathcal{D}(x)\vert dx.
\end{eqnarray}
We see that all quantum interferences between different position detection events have been removed. Furthermore, if the shifts $\pm x_0$ are much larger than the width of the Gaussian wavepackets, they do not overlap spatially on the detector in practice. Then, the first integral provides unambiguously the probability $\vert\alpha_H\vert^2$ of detecting the photon with horizontal polarisation around the location $x_0$, while the second one provides the corresponding probability $\vert\alpha_V\vert^2$ of detecting the photon with vertical polarization around the location $-x_0$.}

{While such an account elucidates why we obtain classical states and provides their respective probabilities, it does not explain why a single outcome emerges. Indeed, the mixed state (\ref{eq:decoherence2}) only reflects our ignorance of the exact environment state. The true quantum state of the full setting remains the giant superposition given by (\ref{eq:decoherence1}), where all the coherences are preserved, although hidden in the environment states and inaccessible in practice. From this point of view, the decoherence approach does not solve the measurement problem \cite{Hance}. Because such a state is still in a superposition and not in any eigenstate of the photon, (\ref{eq:decoherence1}) does not allow us to claim that a property of the photon has been measured, at least not until we apply the projection postulate to select one of the outcomes. In other words, decoherence theory cannot yet get rid of postulate 2 to determine when a system has a property, according to orthodox QM. Therefore, even with a detailed decoherence model accounting for the photon’s destructive interaction with the detector, the paradox still arises if the projection is applied after the photon is destroyed. From a decoherence perspective—which aims to explain classical results without invoking the projection postulate—it makes little sense (in our opinion) to apply the projection before the environment has significantly interacted with the detector, and the detector with the photon. For intellectual consistency, the projection should therefore be placed after the photon has been detected and destroyed. In any case, the projection cannot occur before the photon reaches the detector, since up to that point its evolution and propagation remain perfectly reversible, as demonstrated in many experiments. At this stage, decoherence approaches, while promising, appear unable to deal properly with the paradox described here.}

\section{Conclusion}

When doing a measurement on a microsystem, as a physicist one may have a very strong intuition that the property just measured is a property of an object that was present in the measuring device just prior to measurement. But however strong this intuition may be, when it comes to {the} destructive measurements of properties of photons {discussed in this paper} this intuition cannot be expressed with the formalism and postulates of OQM. Consequently, we arrive at this conclusion: even when destructive experiments on photons are state-of-the-art technologically, their outcomes cannot be cited as support for OQM. Fairly recently, Wheeler called the Copenhagen interpretation ``the best interpretation of the quantum that we have'' \cite{Wheeler}; but as it is, we disagree---this interpretation fails woefully in the treatment of destructive measurements of properties of photons.

\section*{Acknowledgments}
Y. C. is a research associate of the Fund for Scientific Research F.R.S.-FNRS. Y.C. acknowledges support from the European Union’s HORIZON-EIC-2022-PATHFINDERCHALLENGES-01 under the Grant agreement ID: 101115149 (ARTEMIS project, ``Molecular materials for on-chip integrated quantum light sources"). This work reflects only the authors' view, and the European Commission is not responsible for any use that may be made of the information it contains.

\section*{Author Declarations}

\subsection*{Conflict of interest}

The authors have no conflicts to disclose

\subsection*{Data availability}

Data sharing is not applicable to this article as no new data were created or analyzed in this study.


\begin{thebibliography}{0}

\bibitem{Bell}
J. Bell, ``On the Einstein Podolsky Rosen Paradox'', \emph{Physics} \textbf{1}(3), 195–200 (1964)

\bibitem{Freedman}
S.J. Freedman, J.F. Clauser, ``Experimental test of local hidden-variable theories'', \emph{Phys. Rev. Lett.} \textbf{28}(938), 938--941 (1972).

\bibitem{Aspect}
A. Aspect, J. Dalibard, G. Roger, ``Experimental Test of Bell's Inequalities Using Time-Varying Analyzers". \emph{Phys. Rev. Lett.} \textbf{49}(25), 1804--1807  (1982)

\bibitem{Zeilinger}
J. Pan, D. Bouwmeester, D., M. Daniell, H. Weinfurter, A. Zeilinger, ``Experimental test of quantum nonlocality in three-photon GHZ entanglement", \emph{Nature} \textbf{403}(6769), 515–519 (2000)

\bibitem{Muller1}
F.A. Muller, \emph{Quantum Mechanics for Philosophers.} Syllabus. Erasmus University Rotterdam (2025); available from the author on request

\bibitem{Muller2}
F.A. Muller, \emph{Six Measurement Problems of Quantum Mechanics}, in: J.R.B. Arenhart, R.W. Arroyo (eds.), \emph{Non-Reflexive Logics, Non-Individuals, and the Philosophy of Quantum Mechanics.} Springer, pp. 225–260 (2023)

\bibitem{Fine}
A. Fine, ``Probability and the Interpretation of Quantum Mechanics'', \emph{British J. Philos. Sci.} \textbf{24}(1), 1--37 (1973)

\bibitem{Dirac}
P.A.M. Dirac, \emph{Principles of Quantum Mechanics}, Oxford University Press  (1930)

\bibitem{Heisenberg}
{W. Heisenberg, \emph{The Physical Principles of the Quantum Theory}, Dover Publications (1930)}

\bibitem{Muller3}
F.A. Muller, \emph{Circumveiloped by Obscuritads. The nature of interpretation in quantum mechanics, hermeneutic circles and physical reality, with cameos of James Joyce and Jacques Derrida}, in: J.Y. Béziau, D. Krause, J.R.B. Arenhart (eds.), \emph{Conceptual Clarifications. Tributes to P.C. Suppes (1922–2014)}, College Press (2015); arXiv:1406.6284

\bibitem{VonNeumann1}
J. von Neumann, \emph{Mathematical Foundations of Quantum Mechanics} (R.T. Beyer, trans.), Princeton University Press, pp. 347-348 (1955; original work published in 1932)

\bibitem{Luders}
{G. L\"{u}ders, ``\"{U}ber die Zustands\"{u}nderung durch den Me{\ss}proze{\ss}'', \emph{Ann. Phys.} \textbf{443}, 322-328 (1950)}

\bibitem{Sudbery}
A. Sudbery, ``Whose Projection Postulate?'', arXiv:2402.15280 (2024)

\bibitem{Weinberg}
S. Weinberg, ``The greatest physics discovery of the 20th century'', \emph{J. Phys. Conf. Ser.} \textbf{2877}, 012114 (2024)

\bibitem{Braginsky}
V.B. Braginsky, Y.I. Vorontzov, K.S. Thorne, ``Quantum nondemolition measurements'', \emph{Science} \textbf{209}(4456), 547–557 (1980)

\bibitem{Haroche}
G. Nogues, A. Rauschenbeutel, S. Osnaghi, M. Brune, J.M. Raimond, S. Haroche, ``Seeing a single photon without destroying it'', \emph{Nature} \textbf{400}(6741), 239--242 (1999)

\bibitem{Cabbolet2}
M.J.T.F. Cabbolet, ``A category mistake in observational claims regarding ultrashort-lived unstable particles'', \emph{Mod. Phys. Lett. A} \textbf{33}(38), 1850227 (2018)

\bibitem{Nielsen}
{M.A. Nielsen and I.L. Chuang, \emph{Quantum computation and quantum information}, Cambridge University Press (2000)}

\bibitem{Barnett}
{S.M. Barnett and S. Croke, ``Quantum state discrimination", \emph{Adv. Opt. Photon.} \textbf{1}, 238--278 (2009)}

\bibitem{Aharonov}
{Y. Aharonov, D.Z. Albert, and L. Vaidman, ``How the result of a measurement of a component of the spin of a spin-1/2 particle can turn out to be 100", \emph{Phys. Rev. Lett.} \textbf{60}, 1351--1354 (1988)}

\bibitem{Dressel}
{J. Dressel, M. Malik, F. Miatto, A.N. Jordan, R.W. Boyd, ``Colloquium: Understanding quantum weak values: Basics and applications", \emph{Rev. Mod. Phys.} \textbf{86}, 307--316 (2014)}

\bibitem{Hall}
{M. Hall, ``Prior information: how to circumvent the standard joint-measurement uncertainty relation", \emph{Phys. Rev. A}, \textbf{69}, 052113 (2004)}

\bibitem{Dressel2}
{J. Dressel, ``Weak values as interference phenomena", \emph{Phys. Rev. A} \textbf{91}, 032116 (2015)}

\bibitem{Xu}
{L. Xu and L. Zhang, ``Progress and perspectives on weak-value amplification", \emph{Prog. Quantum Electron.} \textbf{96}, 100518 (2024)}

\bibitem{Ballesteros}
{L. Ballesteros Ferraz, J. Martin, and Y. Caudano, ``On the relevance of weak measurements in dissipative quantum systems", \emph{Quantum Sci. Technol.} \textbf{9}, 035029 (2024)}

\bibitem{Zurek1}
{W.H. Zurek, ``Decoherence, einselection, and the quantum origins of the classical", \emph{Rev. Mod. Phys.} \textbf{75} {715--775} (2003)}

\bibitem{Zurek2}
{W.H. Zurek, ``Quantum Darwinism", \emph{Nature Phys.} \textbf{5}, 181--188 (2009)}

\bibitem{Zurek3}
{W.H. Zurek, ``Quantum theory of the classical: quantum jumps, Born’s Rule and objective classical reality via quantum Darwinism", \emph{Phil. Trans. R. Soc. A} \textbf{376}, 20180107 (2018)}

\bibitem{Hance}
{J.R. Hance and S. Hossenfelder, ``What does it take to solve the measurement problem?", \emph{J. Phys. Commun.} \textbf{6}, 102001 (2022)}

\bibitem{Wheeler}
J.A. Wheeler, ``A Practical Tool, But Puzzling, Too'', \emph{New York Times}, Edition December 12 (2000)

\end{thebibliography}
\end{document}